\documentclass[article,superscriptaddress,twocolumn,letterpaper,prb]{revtex4}
\usepackage{fullpage}
\usepackage{graphics}
\usepackage{fullpage}
\usepackage{graphicx}
\usepackage{epstopdf}
\begin{document}
\newtheorem{theorem}{Theorem}
\newtheorem{acknowledgement}[theorem]{Acknowledgment}
\newtheorem{algorithm}[theorem]{Algorithm}
\newtheorem{axiom}[theorem]{Axiom}
\newtheorem{claim}[theorem]{Claim}
\newtheorem{conclusion}[theorem]{Conclusion}
\newtheorem{condition}[theorem]{Condition}
\newtheorem{conjecture}[theorem]{Conjecture}
\newtheorem{corollary}[theorem]{Corollary}
\newtheorem{criterion}[theorem]{Criterion}
\newtheorem{definition}[theorem]{Definition}
\newtheorem{example}[theorem]{Example}
\newtheorem{exercise}[theorem]{Exercise}
\newtheorem{lemma}[theorem]{Lemma}
\newtheorem{notation}[theorem]{Notation}
\newtheorem{problem}[theorem]{Problem}
\newtheorem{proposition}[theorem]{Proposition}
\newtheorem{remark}[theorem]{Remark}
\newtheorem{solution}[theorem]{Solution}
\newtheorem{summary}[theorem]{Summary}
\def\r{{\bf{r}}}
\def\j{{\bf{j}}}
\def\m{{\bf{m}}}
\def\k{{\bf{k}}}
\def\kt{{\tilde{\k}}}
\def\mt{{\hat{t}}}
\def\mG{{\hat{G}}}
\def\mg{{\hat{g}}}
\def\mGa{{\hat{\Gamma}}}
\def\mS{{\hat{\Sigma}}}
\def\mT{{\hat{T}}}
\def\K{{\bf{K}}}
\def\P{{\bf{P}}}
\def\q{{\bf{q}}}
\def\Q{{\bf{Q}}}
\def\p{{\bf{p}}}
\def\x{{\bf{x}}}
\def\X{{\bf{X}}}
\def\Y{{\bf{Y}}}
\def\F{{\bf{F}}}
\def\G{{\bf{G}}}
\def\bG{{\bar{G}}}
\def\mbG{{\hat{\bar{G}}}}
\def\M{{\bf{M}}}
\def\V{\cal V}
\def\tchi{\tilde{\chi}}
\def\tx{\tilde{\bf{x}}}
\def\tk{\tilde{\bf{k}}}
\def\tK{\tilde{\bf{K}}}
\def\tq{\tilde{\bf{q}}}
\def\tQ{\tilde{\bf{Q}}}
\def\si{\sigma}
\def\ep{\epsilon}
\def\hep{{\hat{\epsilon}}}
\def\al{\alpha}
\def\be{\beta}
\def\ep{\epsilon}
\def\bep{\bar{\epsilon}_\K}
\def\up{\uparrow}
\def\de{\delta}
\def\De{\Delta}
\def\up{\uparrow}
\def\dwn{\downarrow}
\def\ksi{\xi}
\def\etha{\eta}
\def\product{\prod}
\def\goto{\rightarrow}
\def\switch{\leftrightarrow}
\def\switch{\leftrightarrow}
\title{Does Singlet Fission Enhance the Performance of Organic Solar Cells?}
\author{K. Aryanpour}
\author{J.~A.~Mu\~noz}
\author{S. Mazumdar \\ {\it Department of Physics, University of Arizona, Tucson, Arizona 85721, United States}}
\date{\today}
\begin{abstract}
\noindent{\bf ABSTRACT:} Singlet fission, in which the optical spin-singlet exciton dissociates into two low energy triplet excitons, has been proposed as a viable approach to enhance the quantum efficiency of organic solar cells. We show that even when singlet fission is occurring in the donor molecule, the electronic structure at the donor$-$acceptor interface must satisfy specific requirements for the solar cell performance to be enhanced by this process. We focus on the pentacene$-$C$_{60}$ solar cell, and on the basis of our calculations and available experimental data, we conclude that there is not enough evidence that these requirements are met by the donor$-$acceptor interface here. We propose experiments that can determine whether the minimal requirement for enhanced performance driven by singlet fission is met in this and other solar cells.
\end{abstract}
\maketitle
\noindent{\bf Introduction} 
\par Multiple exciton generation (MEG), involving the generation of two or more low energy excitons from the absorption of a single high energy photon, \cite{Nozik10a} has been suggested as an important means to overcome the Shockley$-$Queisser upper limit of 33\% \cite{Shockley61a} for the quantum efficiency (QE) of inorganic solar cells. Within the MEG scenario, an optically generated high energy exciton undergoes conversion to several lower energy excitons, while obeying energy conservation. If each of the low energy excitons now undergoes dissociation into electron and hole carriers, the QE will exceed the limit for a single exciton. A related process, singlet fission (SF), has generated considerable excitement in the context of organic solar cells. \cite{Lee09a,Rao10a,Zimmerman10a,Zimmerman11a,Smith10a,Greyson10a,Greyson10b,Johnson10a,Chan11a,Chan12a,Wilson11a,Thorsmolle09a,Burdett10a,Muller06a,Muller07a,Berkelbach12a,Berkelbach12b} In organic $\pi$-conjugated materials, total spin is usually a good quantum number, and exchange interactions are large. The spin selection rule limits optical absorption to spin singlet states only, with most of the oscillator strength concentrated in the lowest optical exciton in the quasi-one-dimensional (quasi-1D) materials. \cite{Barford05a,Mazumdar97a} It has long been recognized that in many $\pi$-conjugated systems, the energies of the optical spin singlet state and the lowest triplet state satisfy the inequality $E_{\mathrm{S}} \geq 2E_\mathrm{T_{1}}$, where $E_\mathrm{S}$ ($E_\mathrm{T_{1}}$) is the singlet (triplet) exciton energy. Indeed, in many such systems, the lowest singlet state is not the optical exciton but an optically forbidden two-photon state that is an entangled state of two triplets (hereafter TT). \cite{Hudson82a,Ramasesha84a,Tavan87a} In principle, such a system is a candidate for SF, whereby the singlet optical exciton undergoes fission into two triplet excitons. While SF in organic materials has been known for a long time, \cite{Pope99a} recent excitement began with the observation of relatively high power conversion efficiency of organic solar cells with pentacene (hereafter PEN) as the donor (D) molecule and C$_{60}$ as the acceptor (A). \cite{Yoo04a,Yoo07a} Experimental demonstrations of SF in tetracene and pentacene crystals \cite{Lee09a,Rao10a,Wilson11a,Geacintov69a,Groff70a,Burgos77a,Jundt95a,Sakamoto04a,Thorsmolle09a,Burdett10a,Muller06a,Muller07a,Johnson09a,Singh10a} have led to the idea that the enhanced performance of PEN$-$C$_{60}$ solar cells is due to SF.
\begin{figure}
\includegraphics[width=2.2in]{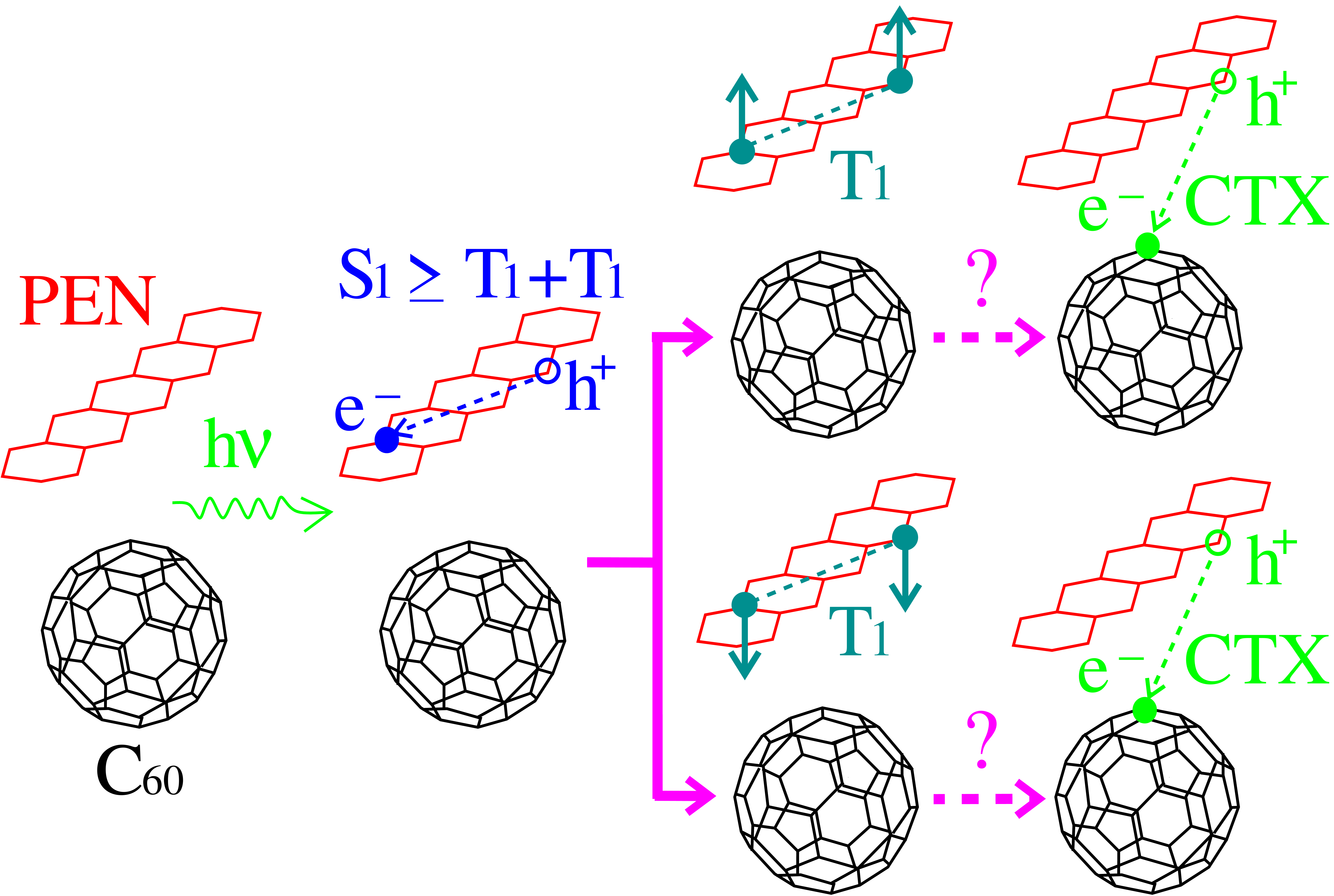}\\
\vspace{0.1in}
Table of Contents (TOC) figure.
\label{TOC}
\end{figure} 
\par With few exceptions, \cite{Lee09a,Chan11a,Rao10a} research on SF has been limited to single-component systems, with the focus on the determination of the mechanism of SF (in particular, in the acenes). \cite{Zimmerman10a,Zimmerman11a,Smith10a,Greyson10a,Greyson10b,Pope99a,Geacintov69a,Groff70a,Burgos77a,Berkelbach12a,Berkelbach12b,Jundt95a,Sakamoto04a,Thorsmolle09a,Johnson09a,Burdett10a,Muller06a,Muller07a} Whether or not SF can give enhanced performance, however, depends on the electronic structures of both D and A, and in particular, of the D$-$A interface. The goal of our work is different from the bulk of the existing theoretical work on SF \cite{Zimmerman10a,Zimmerman11a,Smith10a,Greyson10a,Greyson10b} and is complementary to this research; it is to determine {\it the conditions which need to be satisfied for SF-driven higher QE in organic solar cells.} 
\par At the heart of organic photovoltaics lies photoinduced charge-transfer (PICT) between D and A at their interface. Optically excited D (A) donates an electron (hole) to its partner, generating a charge-transfer exciton D$^{+}$A$^{-}$. The lowest energy charge-transfer exciton (hereafter CTX$_0$), depending upon its binding energy, now undergoes recombination as well as charge-separation, and only the latter process is useful in photovoltaics. For SF to give enhanced photovoltaic performance, each of the two molecular triplets should now donate an electron or a hole to its partner. One then sees right away that for SF-driven higher performance each of the following conditions have to be satisfied: (i) $E_\mathrm{CTX_{0}} \leq E_\mathrm{T_1}$, where $E_\mathrm{CTX_{0}}$ is the energy of the lowest charge-transfer exciton, (ii) the binding energy of CTX$_0$ should not be prohibitively large, and (iii) the ground state should continue to be neutral covalent and should not have undergone transition to an ionic state (such neutral-to-ionic transition, for example, occurs in crystalline mixed-stack charge-transfer solids \cite{Torrance81a}). We have not attempted to distinguish between spin-singlet and triplet CTX$_0$ in the above. Because of the large electron$-$hole separation, singlet and triplet CTX$_0$ 
are known to be nearly degenerate, as has been shown previously \cite{Kadashchuk04a} and as we have confirmed in our calculations. Note that condition i requires a very low energy CTX$_0$, given that $E_\mathrm{T_1}$ should satisfy $E_\mathrm{S} \geq 2E_\mathrm{T_1}$. Indeed, such low energy triplets are covalent in the valence bond language, \cite{Ramasesha84a,Tavan87a} suggesting their unsuitability in charge-transfer processes. Conditions ii and iii have to be therefore satisfied despite the very low $E_\mathrm{CTX_0}$. 
\par In the work presented here, we report explicit calculations of PICT on quasi-1D and quasi-two-dimensional (quasi-2D) PEN$-$C$_{60}$ systems. We show from comparisons of detailed calculations and available experimental information that whether or not SF in PEN is expected to give higher performance of the PEN$-$C$_{60}$ solar cell is not necessarily obvious. Even if SF-driven enhanced performance is occurring in this system, we show that PEN$-$C$_{60}$ is a marginal case where the above conditions are barely satisfied, and it cannot be assumed that all the molecules \cite{Smith10a} that are being investigated as candidates for SF will necessarily lead to higher QE for PICT. Elaborate evaluations of the above conditions for each D$-$A pair are essential because, for systems in which they are not satisfied, SF provides a competing channel for the decay of the photoexcitation and diminished performance. 
\par In the next section we present our theoretical model and discuss the computational approach we have taken to simulate the PEN$-$C$_{60}$ interface and to determine the binding energy of CTX$_0$ for idealized 1D and 2D cases. Following this, we present our results for parameters appropriate for isolated PEN and C$_{60}$ molecules, as well as for the solid-state heterostructure. Our calculations reproduce the known experimental results for isolated PEN and C$_{60}$ molecules almost quantitatively. This is because our parametrization of the PPP Hamiltonian has been performed with considerable care. \cite{Chandross97a} Our calculated absolute exciton energies as well as the exciton binding energies in single-walled carbon nanotubes, \cite{Wang06a} and in one- and two-photon states in polycyclic hydrocarbons that are molecular fragments of graphene, \cite{Aryanpour12a} have shown similar quantitative agreement between theory and experiment. We recognize that the agreements with molecular data do not prove that the theory will give quantitatively correct results for intermolecular charge-transfer states. As pointed out at the end of the following section, however, our goal is to merely determine the functional form of the dependence of the energy of CTX$_0$ and its binding energy on the offsets between the molecular orbitals (MOs) of PEN and C$_{60}$. Comparison with experimentally determined MO offsets now can reveal whether or not SF-mediated charge generation is readily feasible.
\\ \\
\noindent {\bf Theoretical Model and Methods.} 
\par
\begin{figure}
\includegraphics[width=3.2in]{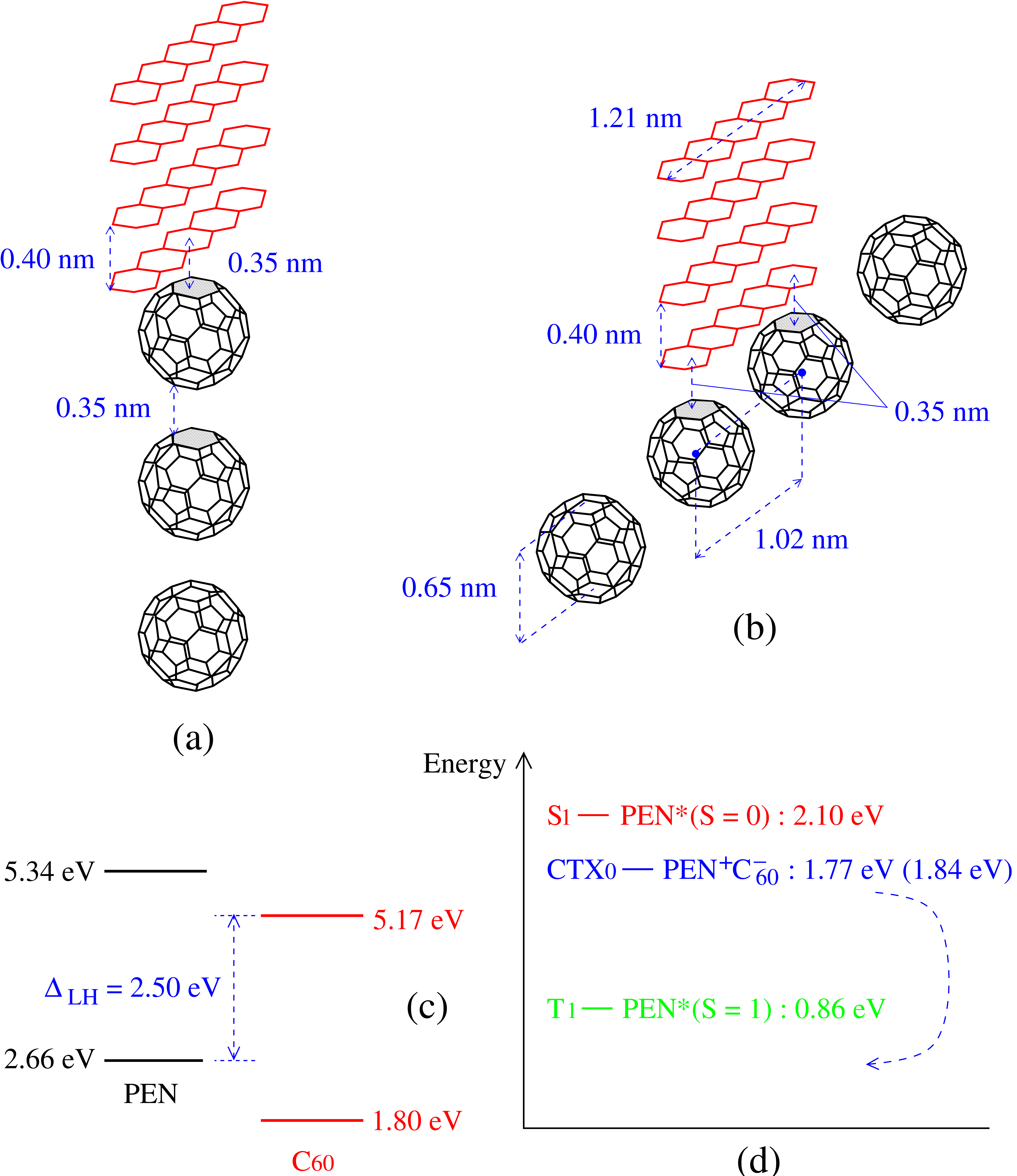}
\caption{Schematics of the PEN$-$C$_{60}$ ``heterostructure'' investigated in our work (see text), in (a) 1D and (b) 2D. Intermolecular separations are as indicated in the figures. (c) The PPP HOMO and LUMO energies of PEN and C$_{60}$ within the HF approximation, and with our Coulomb interaction parameters; the C$_{60}$ HOMOs and LUMOs are five and 3-fold degenerate, respectively. (d) Calculated energies of the singlet optical exciton in PEN and the lowest PEN$^+$C$_{60}^-$ charge-transfer exciton in 1D and 2D (the number in parentheses corresponds to 2D), relative to the ground state, using parameters appropriate for isolated PEN and C$_{60}$, and identical screening parameters for intra- and intermolecular Coulomb interactions. The energy of the triplet T$_1$ is from experiment. \cite{Burgos77a} For SF to give enhanced solar cell efficiency, the lowest charge-transfer exciton should occur below T$_1$, as the curved arrow in the figure indicates.}
\label{f1}
\end{figure}

 We have performed correlated-electron calculations for idealized 1D and 2D heterostructures. The 1D structure we consider consists of four PEN and three C$_{60}$ molecules (see Fig.~\ref{f1}a). We assume the PEN molecules to lie directly above or below one another with an eclipsed geometry. We assume similarly that one hexagonal face of each C$_{60}$ faces a hexagonal face of the next C$_{60}$ molecule, with all intermolecular carbon atoms perfectly aligned. Finally, a hexagonal face of the top C$_{60}$ molecule is taken to be perfectly aligned with the central benzene nucleus of the proximate PEN molecule (see Fig.~\ref{f1}a). The minimum separations between two C$_{60}$ molecules as well as that between the closest PEN and C$_{60}$ are $0.35$ nm in our calculations, while those between the PEN molecules are taken to be $0.40$ nm. These separations are representative of intermolecular separations in organic molecular crystals (including, in particular, epitaxially grown films of C$_{60}$ on VSe$_2$ \cite{Schwedhelm98a}). The 2D structure we have considered consists of four PEN and four C$_{60}$ molecules, as shown in Fig.~\ref{f1}b. Here the PEN and the C$_{60}$ stacks are indvidually 1D but form a ``T-junction'' together. We are aware that the relative orientations between PEN and C$_{60}$ are quite different in the real systems \cite{Verlaak09a,Beljonne11a} but have deliberately chosen these idealized conformations because they will promote maximally stable charge-transfer exction and the most efficient charge separation due to the large intermolecular hoppings that result from the idealized geometry. While recombination is also higher with this geometry, \cite{Yi09a} this is not of concern here. Also, the 2D structure of Fig.~\ref{f1}b, with 328 carbon atoms, is at the limit of our computational capability.
\par Our calculations are within the Pariser-Parr-Pople (PPP) Hamiltonian \cite{Pariser53a,Pople53a} for a two-component system, \cite{Aryanpour10a}
\begin{equation}
\label{Ham}
H = H_{intra}+H_{inter} 
\end{equation}
The intramolecular component $H_{intra}$ is given by,
\begin{eqnarray}
\label{intra}
H_{intra}=-\sum_{\mu\langle ij \rangle, \sigma}t_{ij}^{\mu}
(c_{\mu,i,\sigma}^\dagger c_{\mu,j,\sigma}+ HC) &+& \nonumber \\ 
U\sum_{\mu,i} n_{\mu,i,\uparrow} n_{\mu,i,\downarrow}
 + \sum_{\mu,i<j} V_{ij} (n_{\mu,i}-1)(n_{\mu,j}-1) &-&\nonumber \\ 
\sum_{\mu,i,\sigma}{^{'}}\epsilon_{\mu}n_{\mu,i,\sigma} \hspace{1.9in}
\end{eqnarray}
where $c^{\dagger}_{\mu,i,\sigma}$ creates a $\pi$-electron of spin $\sigma$ on carbon atom $i$ of molecule $\mu$, with $\mu=$ 1$-$4 corresponding to PEN, and $\mu=$ 5$-$7 (5$-$8) corresponding to C$_{60}$ in 1D (2D), respectively. HC is Hermitian conjugate, $n_{\mu,i,\sigma} = c^{\dagger}_{\mu,i,\sigma}c_{\mu,i,\sigma}$ is the number of electrons on atom $i$ of molecule $\mu$ with spin $\sigma$ and $n_{\mu,i} = \sum_{\sigma}n_{\mu,i,\sigma}$ is the total number of electrons on atom $i$ of the molecule. $U$ and $V_{ij}$ are the on-site and intramolecular intersite Coulomb interactions, respectively. $V_{ij}$ is obtained from a modification \cite{Chandross97a} of the Ohno parametrization, \cite{Ohno64a} 
\begin{equation}
\label{Ohno}
V_{ij}=U/\kappa\sqrt{1+0.6117 R_{ij}^2}
\end{equation}
where $R_{ij}$ is the distance between carbon atoms $i$ and $j$ in $\mathring{\textrm{A}}$ and $\kappa$ is an effective dielectric constant. Previous work has shown that $U=8$ eV and $\kappa=2$ give excellent fits to absolute exciton energies as well as exciton binding energies in the $\pi$-conjugated polymer poly-paraphenylenevinylene \cite{Chandross97a} and single-walled carbon nanotubes. \cite{Wang06a} We have used standard nearest neighbor one-electron hopping integrals $t_{ij}^{\mu}$ = 2.4 eV for phenyl C-C bonds in PEN, \cite{Chandross97a} and $t_{ij}^{\mu}$ = 1.96 eV and 2.07 eV, respectively, for the bonds within the pentagons of C$_{60}$  and for those linking the pentagons. The smaller C$_{60}$ hopping integrals reflect the curvature that reduces the overlap between neighboring p-orbitals. \cite{Wang06a} We show below that excellent fits to various energy gaps of isolated PEN and C$_{60}$ are obtained with these parameters.
\par The intermolecular component $H_{inter}$ of the Hamiltonian is written as,
\begin{eqnarray}
\label{H_inter}
\label{inter}
H_{inter}=-\sum_{\mu <\mu^{\prime},i,j,\sigma}t^{\perp}_{ij}(c^{\dagger}_{\mu,i, \sigma}
c_{\mu^{\prime},j,\sigma} + HC) \nonumber &+& \\
\frac{1}{2}\sum_{\mu < \mu^{\prime},i,j} V_{ij}^{\perp}(n_{\mu,i} -1)(n_{\mu^{\prime},j} - 1)
\end{eqnarray}
\par We use the same functional form for  $V_{ij}^{\perp}$ as for the intramolecular Coulomb interaction with however a dielectric constant $\kappa_{\perp}$ that may be different from $\kappa$ (smaller $\kappa_{\perp}$ implies stronger intermolecular interaction). Intermolecular hopping integrals follow $t^\perp_{ij}=\beta\exp[(c-d_{ij})/\delta]$, where the prefactor $\beta=$ 0.2 eV, $c$ is the minimum vertical distance between the molecules, $d_{ij}$ is the distance between atom $i$ belonging to molecule $\mu$ and atom $j$ belonging to molecule $\mu^\prime \neq \mu$ and the decay constant $\delta=0.045~\textrm {nm}$. \cite{Uryu04a,Aryanpour10a} 
\par The prime over the summation in the last term in eq.~\ref{intra} indicates that the site energy $\epsilon_{\mu}$ is nonzero only for the atoms of the C$_{60}$ molecules. This term is included to manipulate the energy offsets between the MOs of PEN and C$_{60}$, to vary the energy of CTX$_0$. The energy of the charge-transfer exciton in an arbitrary D$-$A system, in the limit of zero intermolecular hopping, is approximately given by \cite{McConnell65a} IE$_\mathrm{D}$ $-$ EA$_\mathrm{A}$ $-$ $E_\mathrm{C}$, where IE$_\mathrm{D}$ is the ionization energy of D, EA$_\mathrm{A}$ is the electron affinity of A and $E_\mathrm{C}$ is the Coulomb stabilization energy due to proximate oppositely charged D and A in D$^+$A$^-$. This concept in the past has been used mostly for molecular D and A with nondegenerate highest occupied MO (HOMO) and lowest unoccupied MO (LUMO). Very interestingly, we find that it is applicable equally well here to both our 1D and 2D systems, with 5-fold degenerate HOMO and 3-fold degenerate LUMO in C$_{60}$ (see below). IE$_\mathrm{D}$ and EA$_\mathrm{A}$ are both one-electron quantities, and their values in the gas-phase and in the solid-state can differ widely. We simulate the modifications of IE$_\mathrm{D}$ $-$ EA$_\mathrm{A}$ in the solid-state empirically by the effective site energy $\epsilon_{\mu}$, which modulates the mean-field Hartree$-$Fock (HF) energy difference between the LUMO of the acceptor and HOMO of the donor, in accordance with Koopman's theorem. We shall refer to this energy difference as $\Delta_\mathrm{LH}$ in what follows (i.e., $\Delta_\mathrm{LH}=$IE$_\mathrm{D}$ $-$ EA$_\mathrm{A}$). The Coulomb stabilization energy $E_\mathrm{C}$ on the other hand originates from the $V_{ij}^{\perp}$ in eq.~\ref{inter}.
\par Our calculations are mostly within the single configuration interaction (SCI) scheme, including CI between all one electron$-$one hole excitations from the HF ground state, which we take to be a product function of the HF ground states of the individual molecules. This approach enables us to determine ionicities of excited states quantitatively \cite{Aryanpour10a} and also the location of the excited electron and hole in a charge-transfer state. As we show below, this ability to precisely characterize all excited states allows us to determine the binding energy of CTX$_0$ for the model system we are considering. Finally, the SCI assumes the ground state to be neutral and whether or not there is ground state charge-transfer cannot be determined using this approach. We report separate full CI calculations on simpler model systems (see Appendix) to demonstrate that conditions $E_\mathrm{S} \geq 2E_\mathrm{T_1}$ and $E_\mathrm{CTX_{0}} \leq E_\mathrm{T_1}$ can be simultaneously satisfied even with a neutral ground state.  
\par We show below that the PPP-SCI approach reproduces the isolated intramolecular singlet energy states nearly quantitatively. This by itself does not guarantee that our method will also simulate D$-$A interfaces quantitatively, in particular because the parameters $\epsilon_{\mu}$ and $\kappa_{\perp}$ are unknown. Our broad overall conclusions can, however, be arrived at without being able to determine the absolute energy or the binding energy of CTX$_0$ precisely. Our ability to reproduce the energetics of isolated molecules indicates that our calculation of $\Delta_{\mathrm{LH}}$ in the gas-phase is correct. Starting from this limit, we show that $E_{\mathrm{CTX_0}}$ is approximately given by $\Delta_{\mathrm{LH}}-E_{\mathrm{C}}$ for all $\Delta_{\mathrm{LH}}$, and that {\it the same interaction $E_{\mathrm{C}}$ that lowers the energy of CTX$_0$ also raises its binding energy}. Irrespective of whether or not our parametrization of $H_{inter}$, and hence the evaluation of $E_\mathrm{C}$, is correct, it then becomes possible to obtain the correlation between the charge-transfer exciton's binding energy and $\Delta_{\mathrm{LH}}$. From the {\it experimentally} determined values of $\Delta_{\mathrm{LH}}$, one can now estimate whether or not SF-induced enhanced QE in PEN$-$C$_{60}$ is obviously or marginally viable. \\ \\
\noindent {\bf Results} \\ 
\par We begin with computational results for the case where HOMO$-$LUMO offsets between PEN and C$_{60}$ are assumed to be the same as in the gas-phase ($\epsilon_{\mu}=0$) and the screening parameter for the intermolecular Coulomb interactions is the same as for the intramolecular interactions ($\kappa_{\perp}=\kappa$). Following this we simulate solid-state effects by performing calculations for nonzero $\epsilon_{\mu}$ and varying $\kappa_{\perp}$. Finally, we report calculations of the binding energy of CTX$_0$ as a function of $\epsilon_{\mu}$ and $\kappa_{\perp}$.
\\ \\
\noindent \underbar {PEN$-$C$_{60}$ Interface with  ``Gas-Phase'' $\Delta_\mathrm{LH}$.} In Figs.~\ref{f1}c,d we have given our results for parameters appropriate for isolated molecules. Fig.~\ref{f1}c shows the calculated HF HOMOs and LUMOs of PEN and C$_{60}$. Our calculations are within the $\pi$-electron approximation, and the absolute HOMO and LUMO energies are not meaningful. However, all energy {\it differences,} including the calculated $\Delta_\mathrm{LH}$, are relevant. Fig.~\ref{f1}d shows the main results for the PEN$-$C$_{60}$ interface in 1D and 2D, using the MO energies and wave functions corresponding to Fig.~\ref{f1}c and $\kappa_{\perp}=\kappa=2$. Thus, the results of Fig.~\ref{f1}d correspond to the assumption that the molecular HF energies and wave functions are not perturbed at all at the interface in the solid-state. Our calculations are for the singlet states only and the energy of the triplet molecular exciton is taken from experiment. \cite{Burgos77a} 
\par
\begin{figure}
\includegraphics[width=2.9in]{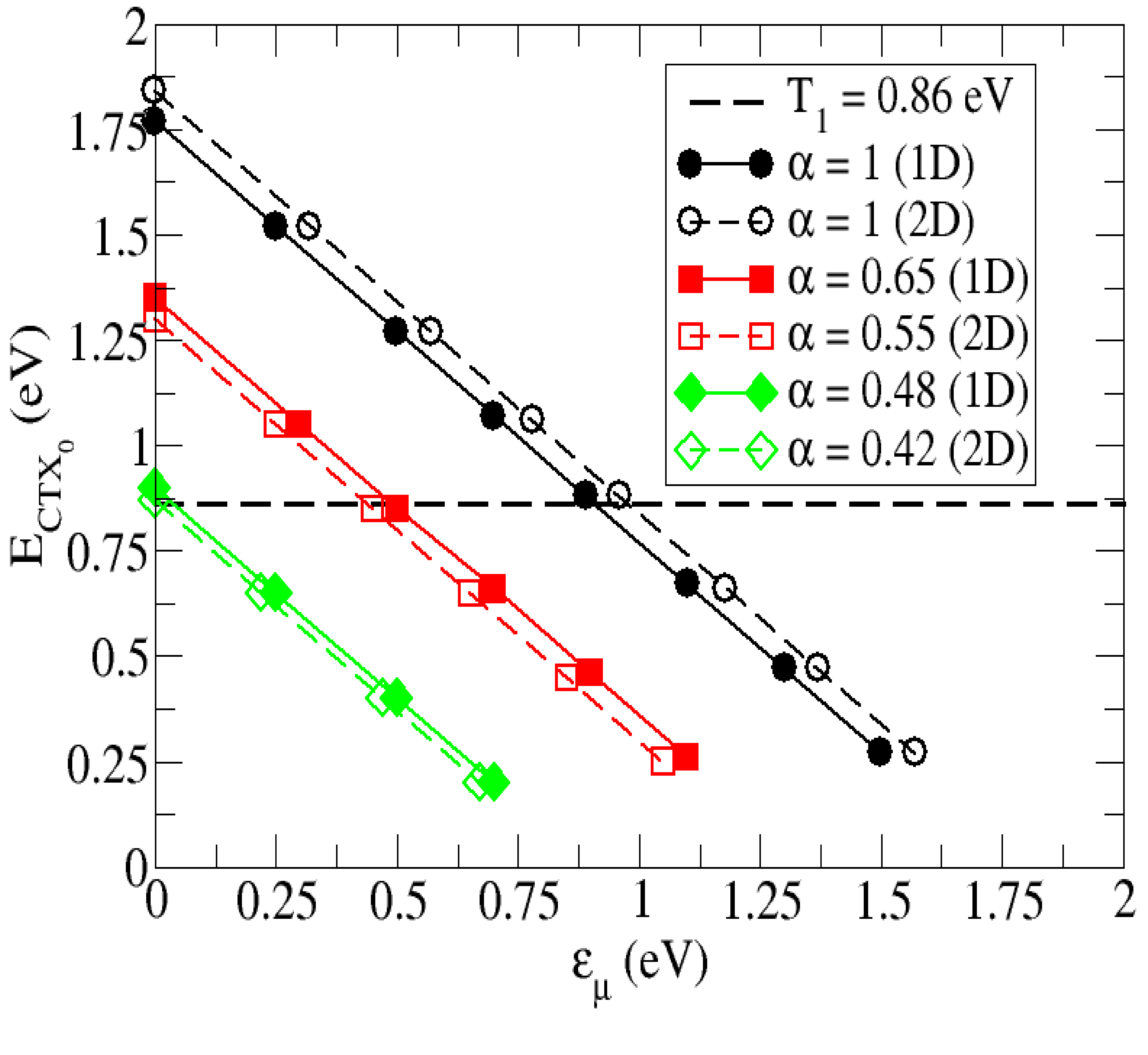}
\caption{$E_\mathrm{CTX_0}$ versus $\epsilon_{\mu}$, the site energies on the carbon atoms of C$_{60}$, in both 1D and 2D for different $\alpha=\kappa_{\perp}/\kappa$ values. The dashed line parallel to the abscissa is the experimental $E_\mathrm{T_1}$ from ref~\onlinecite{Burgos77a}.}
\label{f2}
\end{figure}

 There are several items of interest in the results shown in Figs.~\ref{f1}c,d. Our calculated energy of the optical exciton in PEN, $2.10$ eV, is extremely close to the experimental energy of $2.07$ eV in solution. \cite{Sakamoto04a} Our calculated HOMO$-$LUMO gap for C$_{60}$ $3.37$ eV is practically the same as the experimentally determined value of 3.36 eV. \cite{Schwedhelm98a} The calculated energies for the two lowest allowed optical absorptions in the gas-phase of C$_{60}$, $3.1$ and $3.5$ eV (not shown) are also very close to the experimental values $3.0$ and $3.6$ eV, respectively, for C$_{60}$ molecules dissolved in decalin. \cite{Howard91a} All of these give confidence that our computational approach and the parameters used therein reproduce the behavior of the individual molecules almost quantitatively. Our calculated $E_\mathrm{CTX_0}$ of 1.77 eV in 1D, taken together with the calculated $\Delta_\mathrm{LH}=2.50$ eV, indicate that $E_\mathrm{C}$ is close to $0.7$ eV, provided that the approximate expression $E_\mathrm{CTX_0}=\Delta_\mathrm{LH}-E_\mathrm{C}$ is valid here. Similarly the calculated $E_\mathrm{CTX_0}$ of 1.84 eV in 2D suggests that $E_\mathrm{C}$ is close to 0.66 eV. These values of $E_\mathrm{C}$ are slightly {\it larger} than that calculated from electrostatic considerations for PEN$^+$C$_{60}^-$ at an intermolecular distance of $0.35$ nm (see Fig.~5 in ref~\onlinecite{Yi09a}). Had we evaluated $E_\mathrm{CTX_0}$ from IE$_\mathrm{D}$ $-$ EA$_\mathrm{A}$ $-$ $E_\mathrm{C}$, using the known ``bare'' ionization energy of PEN $6.59$ eV \cite{Gruhn02} and bare electron affinity of C$_{60}$ $2.68$ eV, \cite{Wang05a} along with the calculated $E_\mathrm{C}$ of ref~\onlinecite{Yi09a}, the calculated $E_{\mathrm{CTX_0}}$ would have been significantly larger than those obtained by us. It is then reasonable to hold our calculated $E_\mathrm{CTX_0}$ as a realistic {\it lower limit} for the energy of the charge-transfer exciton with gas-phase parameters. As indicated in Fig.~\ref{f1}d, enhanced QE due to SF is not expected here. The curved broken arrow indicates the extent to which the energy of CTX$_0$ needs to be lowered for SF to give higher QE. Smaller $\Delta_\mathrm{LH}$ and/or larger $E_\mathrm{C}$ than  in the gas-phase would be necessary for this. \\ \\
\noindent \underbar {Simulation of Solid-State Effects.} Smaller $\Delta_\mathrm{LH}$ in the solid-state is a consequence of the smaller ionization energy of PEN and the larger electron affinity of C$_{60}$ in the solid-state. As mentioned in the previous section, we simulate solid-state effects phenomenologically by varying $\epsilon_{\mu}$. In Fig.~\ref{f2} we show the calculated $E_\mathrm{CTX_0}$ in 1D and 2D, as functions of $\epsilon_{\mu}$ for several different $\kappa_{\perp}$ that enter into the calculations of $V_{ij}^{\perp}$. It is useful to define and work with $\alpha=\kappa_{\perp}/\kappa$ as a measure of the inter- over intramolecular Coulomb interaction screening strengths throughout our entire computational results. The linear variations of $E_\mathrm{CTX_0}$ with $\epsilon_{\mu}$ are surprising, as they indicate that the simple expression $E_\mathrm{CTX_0} \simeq$ IE$_\mathrm{D}$ $-$ EA$_\mathrm{A}$ $-$ $E_\mathrm{C}$ continues to hold for a broad range of $\epsilon_{\mu}$ and $\alpha$ (with constant $\kappa=2$ and varying $\kappa_{\perp}$), in 1D as well as 2D, even for degenerate HOMO and LUMO in C$_{60}$ and with nonzero electron hoppings between PEN and C$_{60}$. It is conceivable that this is unique to C$_{60}$ as the acceptor because there can be few direct C$-$C intermolecular hoppings in this case. For each $\alpha$ there exists a critical site energy $\epsilon_{\mu}^c$ below which $E_\mathrm{CTX_0} < E_\mathrm{T_1}$, in both 1D and 2D.
\begin{figure*}
\includegraphics[width=6.6in]{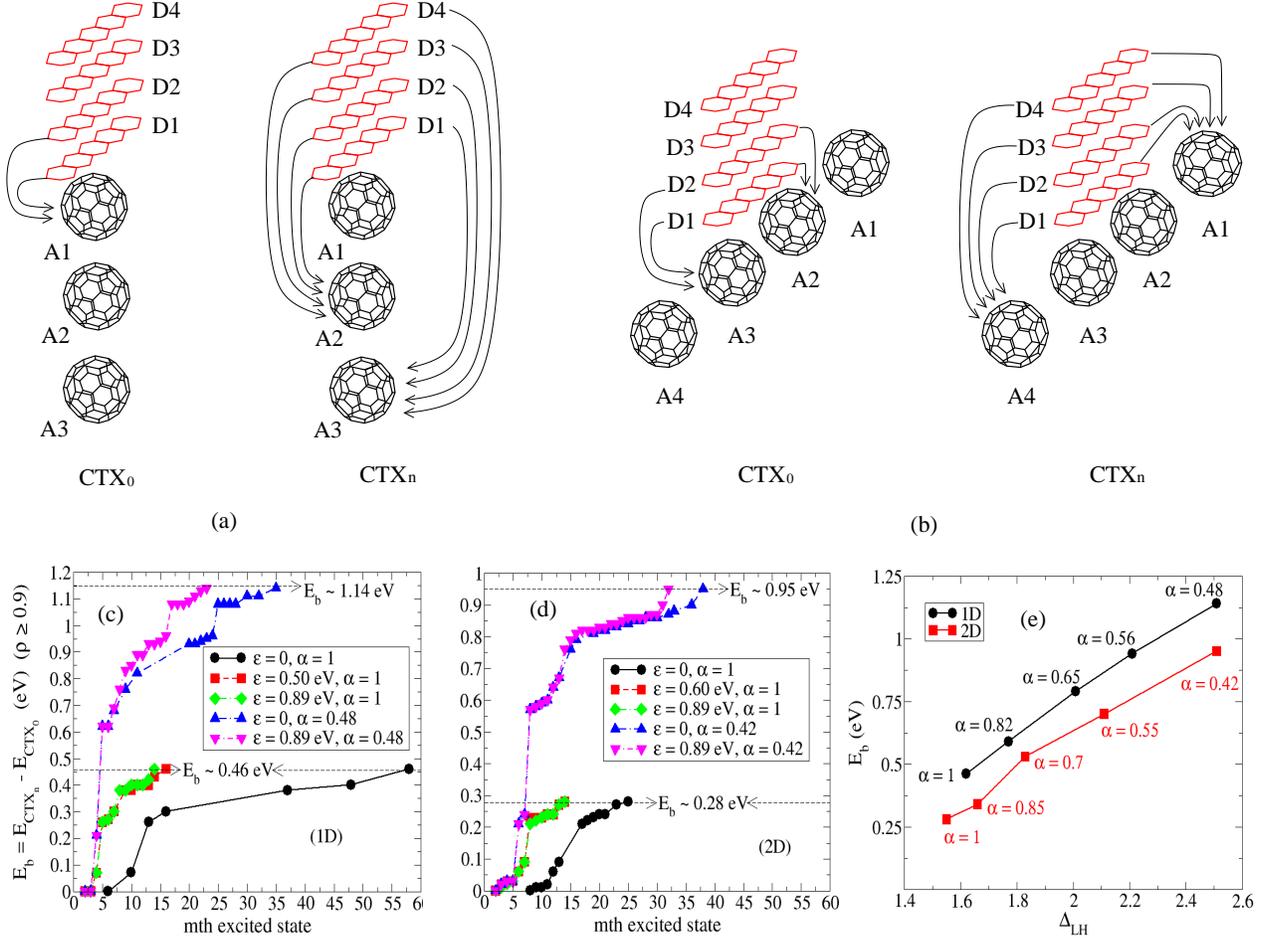}
\caption{(a) Schematic of the lowest energy charge-transfer exciton CTX$_0$ and the high energy charge-transfer exciton CTX$_n$ in 1D (D and A indicate donor and acceptor, respectively). The arrows denote the lengths of the charge-transfer ``bonds''. The wave function of CTX$_0$ is predominantly a superposition of the configuration with the hole on the PEN molecule and the electron on the C$_{60}$ molecule at the PEN$-$C$_{60}$ interface and the configuration with the hole on the neighboring PEN molecule while in CTX$_n$ the hole and the electron are significantly more delocalized, as far away as the farthest PEN and C$_{60}$ molecules from the interface, respectively. (b) CTX$_0$ and CTX$_n$ with their associated charge-transfer schematics in 2D. (c and d) Energy differences between all higher energy charge-transfer excitons and CTX$_0$, versus the quantum numbers of the higher excited states in 1D and 2D, respectively. The terminal point on each plot is CTX$_n$ of parts a and b. (e) $E_\mathrm{b}$ against $\Delta_\mathrm{LH}$, each with $\alpha$ values required to satisfy $E_{\mathrm{CTX_0}} < E_\mathrm{T_1}$, in 1D and 2D.}
\label{f3}
\end{figure*}
\par The linear variations of $E_\mathrm{CTX_0}$ in Fig.~\ref{f2} indicate that $\Delta_\mathrm{LH}$ also varies linearly with $\epsilon_{\mu}$. From the calculated $\epsilon_{\mu}^c = 0.89$ ($0.96$) eV for $\alpha=1$ in 1D (2D) (Fig.~\ref{f2}) we find that the critical $\Delta_\mathrm{LH}$ below which $E_\mathrm{CTX_0} < E_\mathrm{T_1}$ is $\sim 1.6$ (1.5) eV for the case where intra- and intermolecular screenings are comparable. We conclude that the lowest PEN$^+$C$_{60}^-$ charge-transfer exciton will be above the lowest triplet exciton unless at least one of the following two conditions are met in 1D (2D) (i) $\Delta_\mathrm{LH}<1.6$ (1.5) eV (ii) $|E_\mathrm{C}|>0.7$ (0.6) eV. \\ \\
\noindent \underbar {Binding Energy of the Lowest Charge-Transfer Exciton.} Our definition of the binding energy of CTX$_0$, $E_\mathrm{b}$, is the usual one: it is the energy difference between CTX$_0$ and the lowest state in which the hole on the PEN molecules and the electron on the C$_{60}$ molecules are free. A relevant question to ask therefore is whether the very low energy of the CTX$_0$ that satisfies $E_{\mathrm{CTX_0}} < E_\mathrm{T_1}$ implies also a large binding energy. If true, this would also impact solar cell performance negatively. We have calculated $E_\mathrm{b}$ for our model heterostructure as a function of $\epsilon_{\mu}$ to address this question.
\par The calculation of $E_\mathrm{b}$ is nontrivial, because despite the multiple PEN and C$_{60}$ molecules retained in our calculations, the overall systems of Figs.~\ref{f1}a,b are still discrete and there is no true continuum. Thus, the binding energy cannot be determined from energy considerations. We determine the threshold of the continuum from wave function analysis. We identify a specific high energy charge-transfer exciton with widely separated and delocalized electron and hole as the lower threshold of the continuum and calculate the energy difference between this state and CTX$_0$ as the lower bound for $E_\mathrm{b}$. The wave function analysis is however complicated because in addition to states with complete charge-transfer there occur in the same energy range many other excited states including neutral C$_{60}$ optically dark states and states with incomplete charge-transfer. \cite{Aryanpour10a} We ignore these additional irrelevant states and consider only excited states with at least 90\% charge-transfer. 
\par In Figs.~\ref{f3}a,b we show schematics of CTX$_0$ and the high energy charge-transfer exciton, hereafter CTX$_\mathrm{n}$, that we use in our evaluation of $E_\mathrm{b}$ in 1D and 2D, respectively. The arrows indicate charge-transfer from the PEN molecules to the C$_{60}$ molecules identified in the figure. In both 1D and 2D, the hole in CTX$_0$ is delocalized over the lower two PEN molecules, with the electron on the nearest C$_{60}$ molecules to the PEN$-$C$_{60}$ interface. The hole and the electron are both delocalized in CTX$_\mathrm{n}$ which has relatively weak contribution from configurations with the electron on the C$_{60}$ nearest to the PEN$-$C$_{60}$ interface. {\it The wave functions of these excitons are nearly independent of} $\epsilon_{\mu}$ and $\kappa_{\perp}$. This is because we are probing only the region $\kappa_{\perp}<\kappa$ where the intermolecular screening is smaller than the intramolecular screening, and CTX$_0$ is at its smallest physical dimension already at $\kappa_{\perp}=\kappa$. The absolute energy of CTX$_0$ however decreases for smaller $\epsilon_{\mu}$ and smaller $\kappa_{\perp}$. Figs.~\ref{f3}c,d present the energy difference between higher energy charge-transfer excitons and CTX$_0$ against the quantum number of the former, in 1D and 2D, respectively, for $\alpha=1$ and $\alpha=0.48$ (0.42 in 2D) and for several different $\epsilon_{\mu}$ in each case. The terminal points for all curves correspond to CTX$_\mathrm{n}$. The consequence of smaller $\Delta_\mathrm{LH}$ (larger $|\epsilon_{\mu}|$) is to shift the quantum number of charge-transfer states relative to the ``irrelevant'' neutral states but has no bearing on their energies. 
\par The exciton binding energy $E_\mathrm{b}$ is independent of $\Delta_\mathrm{LH}$ for fixed $\alpha$; however, it increases strongly as $\alpha$ is decreased. This result has important {\it positive} implication for the effect of SF on the solar cell efficiency: if low energy for CTX$_0$ is obtained predominantly because of the reduction in ionization energy of PEN and enhancement of the electron affinity of C$_{60}$ in the PEN$-$C$_{60}$ heterostructure (i.e., smaller $\Delta_\mathrm{LH}$), the binding energy of CTX$_0$ is not affected and continues to be small. Our calculated binding energy for CTX$_0$ ($0.46$ eV, see Fig.~\ref{f3}c) in 1D for $\alpha=1$ is nearly identical to the calculated estimation by Verlaak et al. ($0.438$ eV) from electrostatic considerations for the charge-transfer exciton in the PEN$-$C$_{60}$ heterostructure. \cite{Verlaak09a} The binding energy in real three-dimensional structure is expected to be smaller. It is then interesting that our calculated $E_\mathrm{b}$ for $\alpha=1$ in 2D is 0.28 eV, which is close to the experimental estimates of $0.20-0.25$ eV in related experimental systems. \cite{Kern11a,Gelinas11a} The actual decrease in $\Delta_\mathrm{LH}$ due to the solid-state effects in the real system may not be sufficient to give $E_{\mathrm{CTX_0}} < E_\mathrm{T_1}$ (see next section), in which case the lowering of $E_{\mathrm{CTX_0}}$ has to be driven by larger $|E_\mathrm{C}|$. We have therefore calculated the critical $\alpha$ required to obtain $E_{\mathrm{CTX_0}} < E_\mathrm{T_1}$ for several different $\Delta_\mathrm{LH} > 1.6$ ($1.5$) eV in 1D (2D), and for each critical $\alpha$ we have determined  $E_\mathrm{b}$. Our calculated results for $E_\mathrm{b}$ against $\Delta_\mathrm{LH}$, now for different $\alpha$ values necessary to bring CTX$_0$ below T$_1$, is shown in Fig.~\ref{f3}e, for 1D and 2D. We see that $E_\mathrm{b}$ increases steeply with $\Delta_\mathrm{LH}$ in both 1D and 2D. Linear fits to the plots give  $E_\mathrm{b} \simeq  0.77(0.70)\Delta_\mathrm{LH}-0.77(0.80)$ eV in 1D (2D), while quadratic fits yield $E_\mathrm{b} \simeq 1.40(1.51)\Delta_\mathrm{LH}-0.16(0.19)\Delta_\mathrm{LH}^2-1.4(1.6)$ eV in 1D (2D). The coefficients of the linear and quadratic terms dependent on $\Delta_\mathrm{LH}$ are thus quite close in 1D and 2D, and although there is no proof, we surmise that these coefficients are similar also in 3D. The fundamental reason behind this is that the one-electron bandwidth in C$_{60}$ is small for any geometry. Taken together with the known binding energies in related (albeit different) systems, \cite{Kern11a,Gelinas11a} the rapid increase in $E_\mathrm{b}$ with $\Delta_\mathrm{LH}$ suggests that for experimental $\Delta_\mathrm{LH}>1.6$ eV in PEN$-$C$_{60}$, $E_\mathrm{b}$ becomes prohibitively large for SF giving enhanced QE. 
\par In summary, (i) PPP calculations within the SCI approximation indicate that the approximate expression IE$_\mathrm{D}$ - EA$_\mathrm{A}$ - $E_\mathrm{C}$ for the lowest charge-transfer exciton is remarkably accurate for PEN$-$C$_{60}$ over a broad range of $\epsilon_{\mu}$ and $\kappa_{\perp}$; (ii) low $E_\mathrm{CTX_0}<E_\mathrm{T_1}$ in the solid-state does not imply a high exciton binding energy of the CTX$_0$ exciton if the lowering of its energy was primarily due to the smaller $\Delta_\mathrm{LH}$ in the solid-state; (iii) on the other hand, if the lowering of the CTX$_0$ energy is because of the larger intermolecular Coulomb interaction in the heterostructure, the exciton binding energy will be substantially enhanced. Higher performance of the solar cell in the latter case is not expected, in spite of SF. Reviewing of the available experimental information, in particular of $\Delta_\mathrm{LH}$, thus becomes essential for determining whether or not SF is beneficial. \\ \\
\noindent {\bf Discussion} \\ 
\par As mentioned in Introduction, the bulk of the existing literature on SF is on single-component
 PEN, with the focus on understanding the mechanism of SF. Only three groups have investigated the
 PEN$-$C$_{60}$ heterostructure in the context of SF \cite{Lee09a,Rao10a,Chan11a} and have concluded that
 SF indeed enhances charge generation. We reexamine aspects of these investigations carefully below.
 We point out that (i) the viewpoint that charge generation is enhanced in PEN$-$C$_{60}$ was not arrived
 at by the three groups independently, and that (ii) alternate interpretations of the experimental
 observations in these references are possible, and hence further experimental and theoretical work
 is necessary to establish beyond doubt that SF is indeed behind the relatively large QE of PEN$-$C$_{60}$
 solar cells. \cite{Yoo04a,Yoo07a} 
\par Lee et al. have claimed external QE $>100$\% for photocurrent generation in a PEN$-$C$_{60}$ multilayer
 photodetector and have ascribed this to enhanced {\it internal} QE of 145\% due to SF in PEN. \cite{Lee09a}
 This work is not based on the solar cell configuration, and it is not clear whether it is possible to predict
 solar cell performance from it. The estimation of the photocurrent efficiency was dependent upon various
 assumptions and appears to be model dependent. It is also significantly larger than the estimate of the peak
 external QE of $58\pm4$\% obtained in the solar cell configuration upon excitation at 670 nm, the absorption
 wavelength of PEN, by Yoo et al.\cite{Yoo04a} Lee et al. further demonstrated that in the presence
 of weak magnetic field ($0.4$ T) the photocurrent in the PEN$-$C$_{60}$ photodetector decreases by about 3\%
 when the laser photoexcitation is at the PEN absorption wavelength, while the photocurrent is unaffected
 when the excitation wavelength corresponded to the region where C$_{60}$ exhibits strong absorption.
 The magnetic field dependence was ascribed to the ``reduction in the singlet character of the TT state'' of
 PEN in the presence of magnetic field. This argument is based on the theory for the reverse process of triplet
fusion which can generate singlets ($T_1+T_1\to S_1$) \cite{Johnson70a} and assumes that the effect of the magnetic
field is the same on the forward and backward reactions. Subsequent to this early work, it
 was recognized that there can be multiple origins of magnetic field-induced decrease or
increase of photocurrent. \cite{Frankevich71a} Further experimental and theoretical
work by many different groups have established that magnetic field dependence of photocurrent can also be a signature
of charge generation from excimers or polaron pairs. \cite{Frankevich92a,Tolstov05a,Xu08a} Both increase and decrease of
photocurrent with increasingly weak magnetic field (up to $150$ mT), for example, is found in P3HT polythiophene, \cite{Xu08a} where SF
plays no role whatsoever. We give below an alternate explanation of the magnetic field dependence of the photocurrent
in the PEN$-$C$_{60}$ photodetector involving PEN excimers.
\par Chan et al. \cite{Chan11a} have contradicted the scenario of single electron transfers from individual T$_1$.
 These authors have determined that a quantum mechanical superposition of the PEN singlet exciton and the TT state is generated
 instantaneously
 upon photoexcitation of PEN. The authors refer to this state as a multiexciton state ME, and claim that {\it multiple} electron
 transfers
 occur either from ME or from yet another multiexciton state ME$^{\prime}$ which originates from ME (the authors do not explicitly
 identify
 the natures of ME or ME$^{\prime}$). Chan et al. discarded the possibility of single electron transfer from ME or ME$^{\prime}$
 to C$_{60}$
 (which will imply ``normal'' instead of enhanced QE of charge-generation) not based on their own work, but entirely on the premise
 that the
 earlier claim of external QE $>100$\% by Lee et al. is correct. We show below that experimental observations do not preclude
 single electron
 transfer from ME$^{\prime}$ to C$_{60}$.
\par Rao and Wilson et al.'s interpretation of their ultrafast spectroscopic measurements of charge-generation
 in PEN$-$C$_{60}$ is the usual one involving single electron transfers from each T$_1$. The authors also believe that
 CTX$_0$ is below T$_1$. \cite{Rao10a,Wilson11a} This latter conclusion is based upon an estimate of $4.5$ eV for the
 EA of C$_{60}$ in the solid-state, \cite{Kanai09a} which is larger by $1$ eV compared to other estimates.
 \cite{Schwedhelm98a} Such a large EA for C$_{60}$, taken together with the solid-state ionization energy of PEN ($5.1$ eV)
 \cite{Hwang09a} give a $\Delta_\mathrm{LH}$ sufficiently small that it could have actually rendered a neutral-ionic
 transition in the ground state (see Appendix) possible, with $E_\mathrm{C}$ close to what is calculated by us and others.
 \cite{Yi09a} The idea of SF-mediated higher QE is based on delayed charge generation in PEN$-$C$_{60}$, the delay being
 2$-$10 ns after photoexcitation. \cite{Rao10a} The delayed charge-generation is ascribed to longer triplet lifetime and
 diffusion time to the heterojunction interface. Note, however, that delayed charge-generation can be from other competing
 long-lived photoexcitations. For instance, a photoinduced absorption (PA) that appears in 85 fs after photoexcitation of PEN
 is assigned to T$_1$ by Wilson et al.\cite{Wilson11a} but to the ME superposition by Chan et al. \cite{Chan11a}
 Very recently, problems associated with identifying the T$_1$ state from ultrafast (instead of continuous wave) PA, in a
 different material expected to exhibit SF, have been pointed out, \cite{Bange2012a} giving indirect support to the viewpoint
 of Chan et al. \cite{Chan11a}
\par We point out that it is possible to explain the peculiarities mentioned above within a scenario that does not
 involve SF (but does involve the TT state). We speculate that the ME$^{\prime}$ state of Chan et al. \cite{Chan11a}
 is a quantum mechanical superposition of a PEN excimer \cite{Marciniak07a,Marciniak09a} and a TT state.
 This interpretation is not very different from that of Chan et al. \cite{Chan11a} The excimer state is itself
 a superposition of the PEN molecular exciton and the spin singlet polaron-pair. \cite{Aryanpour11a} Such a superposition of the
 excimer and the TT is to be expected based on the observation that the optical state, the polaron-pair, and the TT
 are linked through the hopping term of the intermolecular Hamiltonian (eq.~\ref{H_inter}). \cite{Greyson10a} One
 can then hypothesize that single electron transfer to C$_{60}$ occurs from the excimer component of ME$^{\prime}$, while the
 ultrafast PA assigned to T$_1$ by Wilson et al. \cite{Wilson11a} is from the TT component; the latter would be in
 agreement with the observation of Chan et al. \cite{Chan11a} The long lifetimes of the excimer and the TT and their
 large effective masses (and hence slow diffusion times) would contribute to delayed charge generation from ME$^{\prime}$ within
 this scenario. The idea that different components of intermolecular states can separately exhibit their distinct features has
 been demonstrated recently: distinct PAs from different components of the excimer in ordered PPV polymers have been identified
 both theoretically and experimentally. \cite{Aryanpour11a} An alternate interpretation of the magnetic field dependence
 \cite{Lee09a} is now obtained. In the presence of a weak magnetic field the concentration of singlet PEN polaron-pairs
(which are degenerate with the triplet polaron-pairs \cite{Kadashchuk04a}) will decrease. \cite{Frankevich92a,Tolstov05a,Xu08a}
This in turn will reduce the concentration of the excimers
(which are necessarily singlet, \cite{Aryanpour11a} because the molecular triplet is
considerably below the triplet polaron-pair in energy), leading to smaller charge-transfer to C$_{60}$. There can be 
additional contributions to decreased photocurrent from a variety of mechanisms involving excimers and polaron-pair. \cite{Xu08a}
The important point is that the current experiments do not allow interpretation strictly from the SF perspective.
\par Our goal in the above was {\it not} to claim that SF-mediated enhanced charge generation is not occurring in PEN$-$C$_{60}$
 but to simply point out that despite the popularity of this idea, there are reasons to be cautious. On the basis of our
 computational results, we have concluded that the upper limit for $\Delta_\mathrm{LH}$ that can give SF-enhanced
 performance is $1.6$ eV. This is justified by the demonstration that $E_\mathrm{b}$ rises sharply for
 larger $\Delta_\mathrm{LH}$. We reemphasize that while there is no particular justification of our choices
 for $\kappa_{\perp}$, $E_\mathrm{C}$ as calculated by others \cite{Yi09a} is smaller than that found by us,
 which would require even smaller $\Delta_\mathrm{LH}$! 
\par It has been demonstrated that the energy mismatch between levels in heterostructures cannot be obtained from
 studies of the individual semiconductors. \cite{Beljonne11a} It then becomes necessary to examine works that have probed
 the PEN-C${60}$ heterojunction itself and not PEN and C$_{60}$ separately. We are aware of only two such references.
 \cite{Kang06a,Salzmann08a} Kang et al. \cite{Kang06a} find $\Delta_\mathrm{LH}$ of $1.56$ and $1.50$ eV, respectively,
 for C$_{60}$ deposited on PEN (with gold as the substrate), and PEN deposited on C$_{60}$ (also with gold as the sub-strate).
 Thus, in both cases the measured $\Delta_\mathrm{LH}$ is considerably larger than the $0.5$ eV assumed in ref~\onlinecite{Rao10a}
 and would barely satisfy the necessary condition for enhanced QE. Significantly different $\Delta_\mathrm{LH}$ are
 found by Salzmann et al., \cite{Salzmann08a} who find this quantity to depend strongly on the processing technique
 used to generate the heterostructure. For layered structures of C$_{60}$ on PEN precovering PEDOT:PSS, the authors find
 $\Delta_\mathrm{LH} \sim 1.15$ eV (see Figure 3a of ref \onlinecite{Salzmann08a}), which would make enhanced QE feasible.
 However, for codeposited films of PEN and C$_{60}$ the authors determine $\Delta_\mathrm{LH} \sim 1.75$ eV (see Figure 3b
 of ref \onlinecite{Salzmann08a}), which would make enhanced QE unlikely. Even more importantly, in both refs
 \onlinecite{Kang06a} and \onlinecite{Salzmann08a} only the HOMO energies of PEN and C$_{60}$ are determined directly from experiments.
 The LUMO energy of C$_{60}$ is estimated from the transport gap. \cite{Mitsumoto96a} The latter approach gives a
 {\it lower limit} for $\Delta_\mathrm{LH}$: recent work has shown that the transport in molecular solids involves {\it inter}molecular
 charge-transfer states that occur {\it below} the molecular LUMO level. \cite{Aryanpour11a} The true $\Delta_\mathrm{LH}$
 can thus be larger than those estimated in these refs \onlinecite{Kang06a} and \onlinecite{Salzmann08a}. Thus, for example,
 while refs \onlinecite{Kang06a} and \onlinecite{Salzmann08a} assume the HOMO$-$LUMO gap of solid C$_{60}$ to be $2.6$ eV,
 the same quantity is estimated to be $3.36$ eV for epitaxial layers of C$_{60}$ on layered VSe$_2$ single crystals by
 Schwedhelm et al., \cite{Schwedhelm98a} and even larger figures had been reported previously by other investigators.
 \cite{Lof92a,Weaver92a,Reihl95a} The HOMO$-$LUMO gap of Schwedhelm et al., taken together with the HOMO offsets between PEN and
 C$_{60}$ determined experimentally in refs \onlinecite{Kang06a} and \onlinecite{Salzmann08a}, would put $\Delta_\mathrm{LH}$
 outside the region that could give enhanced QE.       
\par In summary, we have presented a careful theoretical analysis of the PEN$-$C$_{60}$ interface within the
 correlated $\pi$-electron PPP model. We simulate solid-state effects by independently varying  parameters that change the energy
 separation $\Delta_\mathrm{LH}$ between the LUMO of C$_{60}$ and the HOMO of PEN, and the many-body Coulomb interaction that
 contributes to the binding energy of the PEN$^+$C$_{60}^-$ charge-transfer exciton. On the basis of these calculations, we conclude
 that while it cannot be ruled out that SF is behind the high QE of the PEN$-$C$_{60}$ solar cell, neither is there unqualified
 support for this scenario from currently available experimental information. A variety of experiments have recently
 detected the CTX$_0$ below the optical gap of the donor polymer in organic heterojunctions.
 \cite{Benson-Smith07a,Hallermann08a,Drori08a,Drori10a,Moghe11a} Interestingly, in all such cases CTX$_0$ has been found to occur at
 1.3$-$1.6 eV, viz.,
 significantly above the $0.9$ eV where the CTX$_0$ in PEN$^+$C$_{60}^-$ needs to occur for enhanced QE. It is worth mentioning
 that in ref \onlinecite{Aryanpour10a}, our computed energy of the charge-transfer exciton in PPV-C$_{60}$ (1.7 eV) is
 close to that found in refs \onlinecite{Drori08a} and \onlinecite{Drori10a} (1.55 eV). While PEN is special because of the
 low energy of T$_1$, there is nothing unique about the intermolecular PEN$-$C$_{60}$ interactions that would give ultralow 
CTX$_0$ here. We propose measurements for the direct detection of CTX$_0$ in PEN$-$C$_{60}$ as in the above systems.
 In particular, direct photoexcitation of CTX$_0$ has been possible despite its low oscillator strength. \cite{Drori08a}
 We propose that similar experiments be performed on the PEN$-$C$_{60}$ heterojunctions.  \\ \\
\begin{figure} 
\includegraphics[width=3.1in]{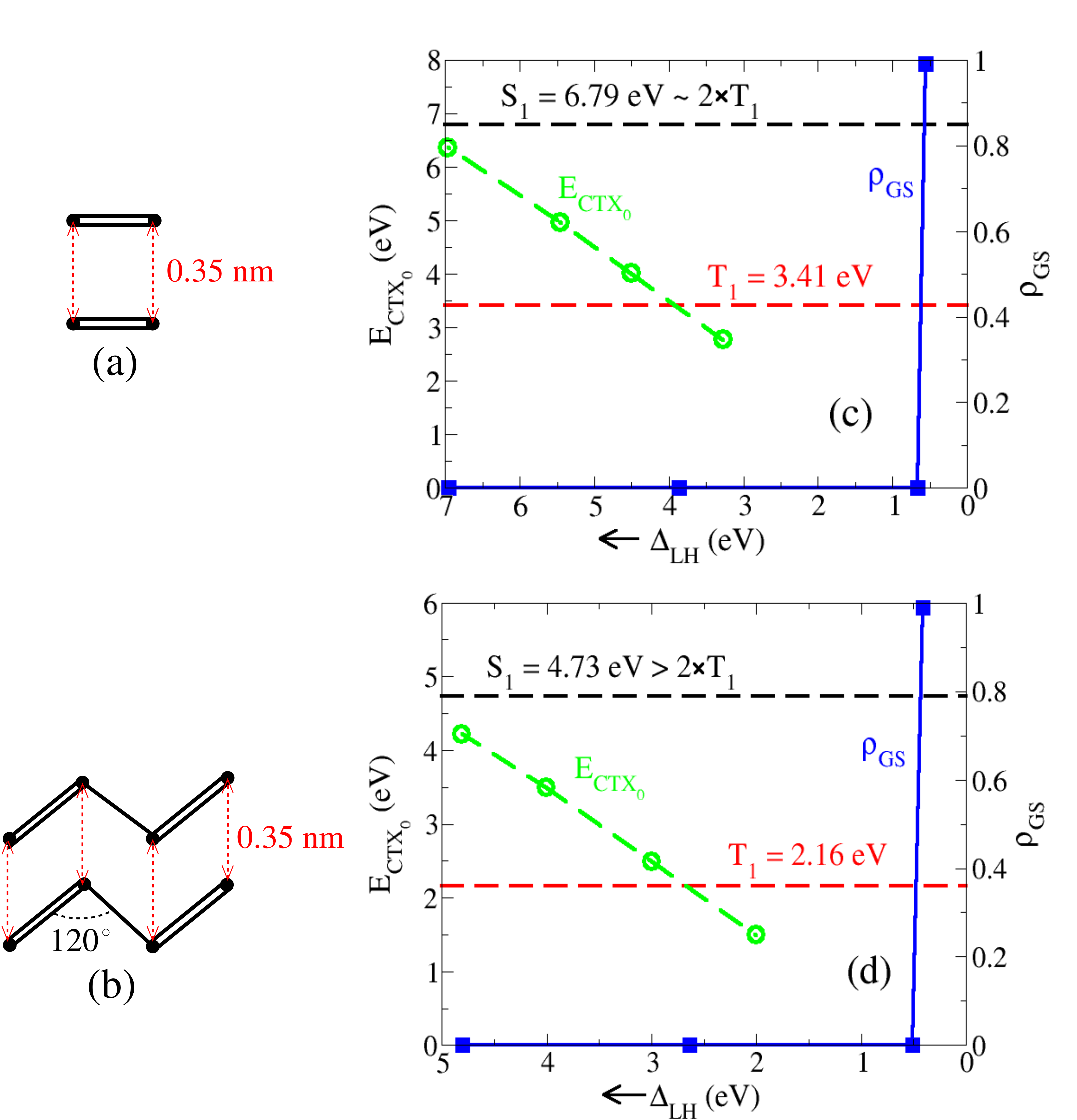}
\caption{(a, b) Interacting pairs of ethylene-like and butadiene-like molecules, with Coulomb parameters chosen such that the lowest molecular triplet is nearly at half the energy of the optical singlet. MO offsets created artificially generate D$-$A systems in both cases. (c, d) $E_\mathrm{CTX_0}$ and the ground state ionicity $\rho_\mathrm{GS}$ versus the MO offset, for the systems in parts a and b, respectively.}
\label{f4} 
\end{figure}
\noindent {\bf Acknowledgment} \\ 
\par We are grateful to Professors Bernard Kippelen (Georgia Tech), Oliver Monti (University of Arizona), Joseph Shinar (Iowa State University), and Zeev Valy Vardeny (University of Utah) for many helpful discussions. We thank Dr. J\'er\^ome Cornil for bringing ref~\onlinecite{Beljonne11a} to our attention. This work was supported by NSF Grant No. CHE-1151475.\\ \\
\noindent {\bf Conflict of Interest} \\ \\
The authors declare no competing financial interest.\\ \\
\noindent {\bf Appendix: Higher Efficiency versus Ground State Charge-Transfer} \\
\par The SCI using HF basis assumes closed shell MOs as the ground state. It is conceivable that as $\Delta_{\mathrm{LH}}$ continues to decrease, this assumption breaks down and the ground state of the D$-$A system becomes ionic, as has been observed in mixed-stack crystalline charge-transfer solids. \cite{Torrance81a} We have considered this possibility because a triplet state that is at half the energy of the singlet exciton is covalent in the valence-bond language, \cite{Ramasesha84a,Tavan87a} and it is conceivable that before the ionic CTX$_0$ can fall below the covalent T$_1$, the ground state itself becomes predominantly ionic, which would have negative consequence for solar cell efficiency. Whether or not this happens cannot be checked within the HF approximation. To test whether the condition $E_{\mathrm{CTX_0}} \leq E_{\mathrm{T_1}}$ can be satisfied without the ground state becoming ionic, we have performed exact diagonalizations of our Hamiltonian for coupled hypothetical small molecules (see Figs.~\ref{f4}a,b). Our systems consist of pairs of molecules that resemble ethylene and butadiene, with the difference that we choose Coulomb parameters such that the triplet exciton is nearly at half the energy of the singlet optical exciton, while keeping all intramolecular and intermolecular electron hoppings realistic. This is achieved by taking $U=5.56$ eV and $\kappa_{\perp}=\kappa=1.7$ (which is not too far from realistic parameters). We include the site energy term in eq.~\ref{intra} for one of the two molecules in each case, thus breaking the symmetry and creating a D$-$A system. We keep varying $\Delta_{\mathrm{LH}}$ slowly and monitor both $E_{\mathrm{CTX_0}}$ and the ground state ionicity (the relative weight of the ionic configuration D$^+$A$^-$ in the wave function) $\rho_\mathrm{GS}$. Our results for the two cases are shown in Figs.~\ref{f4}c,d. In both cases the ground state continues to be covalent where $E_{\mathrm{CTX_0}} \leq E_{\mathrm{T_1}}$ is reached, and the neutral-ionic transition occurs at a much smaller $\Delta_{\mathrm{LH}}$. We have confirmed that this is true for other $\kappa_{\perp}$ also. Essentially, as long as the intermolecular hopping integral has a weak role, as is necessarily true in real D$-$A systems with the C$_{60}$ as the acceptor, $E_{\mathrm{CTX_0}} \leq E_{\mathrm{T_1}}$ can be reached even with covalent ground state.\\ \\
{\it Note added:} During the time this manuscript was going through the review process, a new paper on SF-induced charge generation
in heterostructures with a variety of organic as well as inorganic acceptors and with PEN and diphenyl-pentacene (DPP)
as donors has appeared. \cite{Jadhav12a} The authors list $\Delta_\mathrm{LH}$ for all donor-acceptor pairs, with the frontier orbital
energies obtained from experiments on individual materials, as opposed to in the heterostructures themselves. The
uncertainties in the $\Delta_\mathrm{LH}$ reported by the authors are quite large. Furthermore,
in several cases where $\Delta_\mathrm{LH}$ is smaller than in PEN$-$C$_{60}$ and hence SF-induced charge generation should have been
more likely (such as PEN-diimide and DPP-diimide devices), the external QE are tiny. The authors report density functional theory
calculations for a single PEN$-$C$_{60}$ pair, which find that depending on their relative arrangements the charge-transfer exciton
is either isoenergetic with the molecular triplet of PEN, or occurs at a slightly higher energy. We believe that taken together
these indicate further that even if SF-induced enhanced charge generation occurs in PEN$-$C$_{60}$, this is a marginal case
and each donor-acceptor pair needs to be examined individually.
\end{document}